%% file: paper.tex
\documentclass[sigconf, authorview, 9pt]{acmart}

\PassOptionsToPackage{expansion=false}{microtype}
\usepackage[british]{babel}
\usepackage{graphicx}
\usepackage{booktabs}
\usepackage{xcolor}
\usepackage{tikz}
\usetikzlibrary{positioning,arrows.meta,fit,backgrounds}
\graphicspath{{figures/}}

\usepackage{xcolor}
\usepackage{pgf}
\usepackage{booktabs}
\usepackage{graphicx}

\usepackage{multirow}

\newcommand{\pctchange}[2]{
  \pgfmathsetmacro{\change}{100*(#1-#2)/(#2)}%
  \pgfmathsetmacro{\abschange}{abs(\change)}%
  \ifdim \change pt > 0pt
    {\scriptsize\textcolor{green!60!black}{$\uparrow$\,\pgfmathprintnumber[fixed,precision=0]{\abschange}\%}}%
  \else
    {\scriptsize\textcolor{red!70!black}{$\downarrow$\,\pgfmathprintnumber[fixed,precision=0]{\abschange}\%}}%
  \fi
}

\newcommand{\pctup}[1]{{\scriptsize\textcolor{green!60!black}{(#1)}}}
\newcommand{\pctdown}[1]{{\scriptsize\textcolor{red!70!black}{(#1)}}}

\begin{document}
\microtypesetup{expansion=false}

\setlength{\emergencystretch}{3em}

\title{Towards Traffic Modelling of Multi-Agent Systems: The Role of Coordination Topology}

\author{Davide Lamagna}
\affiliation{%
  \institution{UPC, BarcelonaTech}
  \city{Barcelona}
  \country{Spain}
}
\email{davide.lamagna@estudiantat.upc.edu}

\author{Albert Cabellos}
\affiliation{%
  \institution{UPC, BarcelonaTech}
  \city{Barcelona}
  \country{Spain}
}
\email{acabello@ac.upc.edu}

\author{Alberto Rodriguez-Natal}
\affiliation{%
  \institution{Cisco}
  \city{Barcelona}
  \country{Spain}
}
\email{natal@cisco.com}

\author{G\'abor R\'etv\'ari}
\affiliation{%
  \institution{Budapest University of Technology and Economics}
  \city{Budapest}
  \country{Hungary}
}
\email{retvari@tmit.bme.hu}

\author{Berta Serracanta}
\affiliation{%
  \institution{UPC, BarcelonaTech}
  \city{Barcelona}
  \country{Spain}
}
\email{berta.serracanta@upc.edu}

\renewcommand{\shortauthors}{Lamagna et al.}

\begin{abstract}
Multi-agent LLM systems are an emerging networked workload whose rapid deployment raises questions about the traffic patterns they generate. Compared to conventional applications, these systems generate requests internally: a single user task can induce a structured sequence of model calls whose timing is governed by coordination logic rather than by user arrival rate. It is not clear whether classical traffic models, designed for human-driven workloads, apply to this setting.

We present an empirical characterisation of LLM-call inter-arrival time distributions across sequential, star, and full-mesh agentic coordination topologies, using a multi-layer measurement framework over 500 repeated runs per topology. We find that topology fundamentally shapes the arrival process of requests to the LLM backend: fan-out coordination introduces a structural bimodality absent in sequential execution, and the reasoning-phase component is best described by a log-normal distribution, with the Poisson exponential null model decisively rejected across all topologies. These differences propagate to inference and network level metrics. The framework and analysis pipeline are released openly at \url{https://github.com/dlamagna/agentraffic}.
\end{abstract}



\acmConference[NAIC '26]{Workshop on Networks for AI Computing}{August 17--21, 2026}{Denver, CO, USA}
\acmBooktitle{Workshop on Networks for AI Computing (NAIC '26), August 17--21, 2026, Denver, CO, USA}
\acmDOI{10.1145/3789240.3828749}
\acmISBN{979-8-4007-2467-1/26/08}


\copyrightyear{2026}
\acmYear{2026}
\setcopyright{cc}
\setcctype{by}
\acmConference[SIGCOMM '26]{ACM SIGCOMM 2026 Conference}{August 17--21, 2026}{Denver, CO, USA}
\acmBooktitle{ACM SIGCOMM 2026 Conference (SIGCOMM '26), August 17--21, 2026, Denver, CO, USA}
\acmDOI{10.1145/3789240.3828749}
\acmISBN{979-8-4007-2467-1/2026/08}

\maketitle

\input{body}
\begin{acks}
This work is part of the I+D+i project titled BLOS-SOMS, grant PID2024-158530OB-I00, and is also partially funded by the Catalan Institution for Research and Advanced Studies (ICREA)
\end{acks}

\bibliographystyle{ACM-Reference-Format}
\bibliography{reference}

\end{document}

%% file: body.tex
\section{Introduction}
\label{sec:intro}

Multi-agent large language model (LLM) systems are emerging as a new class of networked application. In these systems, autonomous agents invoke LLMs, call tools, and coordinate with peer agents to complete a task. Unlike conventional human-driven workloads, where traffic is largely driven by external user requests, multi-agent systems can generate additional requests internally as agents reason, branch, retry, and exchange information. A single user task may therefore induce a sequence of model calls and inter-agent messages whose timing is shaped by the coordination logic of the system itself. Parallel branching and coordinated fan-out can produce bursts that matter for shared inference backends, API gateways, service meshes, rate limiters, load balancers, and enterprise networks, where synchronized request arrivals can increase queueing delay even when average request rate is unchanged.

Traffic characterisation has a long history as a tool for planning, dimensioning, and operating communication networks. A recurring lesson from traffic measurement is that new paradigms repeatedly invalidate prior assumptions: Leland et al. showed that Ethernet traffic is self-similar across timescales~\cite{leland1994selfsim}; Paxson and Floyd demonstrated that wide-area traffic frequently violates Poisson assumptions~\cite{paxson1995wide}; Crovella and Bestavros linked Web self-similarity to heavy-tailed transfer sizes~\cite{crovella1997selfsim}; and Benson et al.\ characterised datacenter traffic as bursty, asymmetric, and rack-concentrated: distinct from prior wide-area assumptions~\cite{benson2010datacenter}. History tells us that new workloads (web, peer-to-peer, video streaming, cloud datacenters) often required measurement-driven characterisation before accurate models could be built.

Agentic LLM systems represent a new such workload. Enterprise adoption is already underway: a recent industry survey of over 300 senior executives found that 79\% report AI agents being deployed in their organisations, with 88\% planning to increase budgets specifically because of agentic AI~\cite{pwc2025agents}.
Studying the traffic these systems generate before they become a dominant workload is a natural step in this empirical tradition. To the best of our knowledge, this work is the first to characterise how coordination topology shapes LLM-call arrival processes in multi-agent LLM systems.


In this work, we ask whether the coordination topology of a multi-agent system induces distinct request-arrival processes. A single user-triggered task generates a sequence of internal LLM calls whose timing is shaped by the coordination logic of the agentic workflow, not by user arrival rate. We use the inter-arrival times of these internal calls as our primary traffic metric. We additionally collect lower-layer TCP and packet metrics to provide cross-layer context for the request-level measurements.

A key feature of multi-agent systems is that topology is an explicit design choice~\cite{zhu2025marble,zhang2024gdesigner}: if topology affects how agents solve tasks, it is natural to ask whether it also affects the traffic they generate. We therefore focus on LLM-call inter-arrival times (IATs) across three representative topologies: sequential, star, and full mesh. These topologies capture common coordination patterns: one-at-a-time execution, orchestrator-driven fan-out, and dense peer-to-peer exchange. Our goal is not to exhaust the design space, but to measure how coordination structure is visible in the arrival process.

This paper makes three contributions. First, we provide an initial empirical characterisation of how coordination topology shapes LLM-call IAT distributions.
Second, distribution fitting over the reasoning-phase component shows log-normal is the best fitting model across all topologies, with the exponential null decisively rejected.
Third, we release an open-source, multi-layer measurement framework for studying agentic traffic under configurable coordination topologies. 

\begin{figure}[t]
  \centering
  \includegraphics[width=\columnwidth]{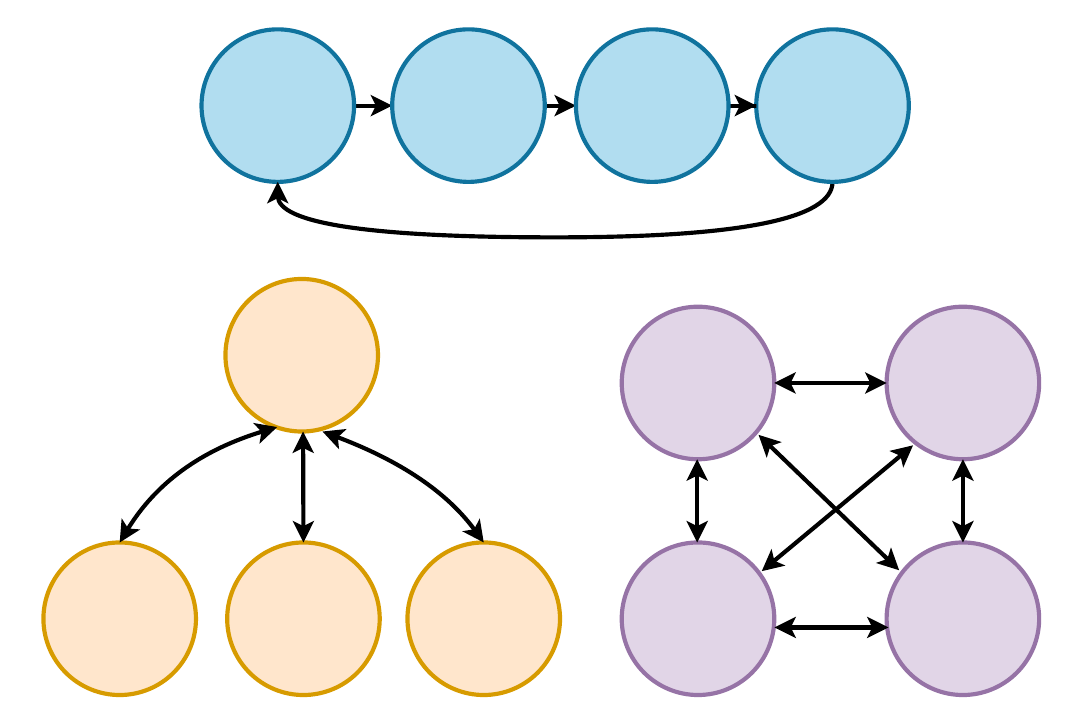}
  \caption{The multi-agent topologies evaluated in this paper: Sequential (top),  Star (bottom left), Full Mesh (bottom right)}
  \label{fig:topologies}
\end{figure}

\section{Background}
\label{sec:background}

\textbf{LLM agents as traffic-generating systems.}
LLM agents differ from conventional request-response applications because they can generate additional work while executing a task. A ReAct-style agent alternates between reasoning and actions, deciding when to query a model, use external sources (tool/agents), or continue execution~\cite{yao2023react}. 
As a result, the arrival process observed by the network can be shaped by internal control flow rather than directly by user arrival rate. Recent work has begun to frame LLM and generative-AI services as an emerging source of Internet traffic ~\cite{koneva2025llmtraffic}; however, multi-agent systems add a further layer of internally generated coordination traffic. 
Agentic LLM systems resemble microservice applications in that a single external request can expand into an internal call graph. However, the graph is not fixed by application code alone: it is also shaped dynamically by agent reasoning, coordination policy, and model or tool outputs. This paper focuses on LLM-call traffic as a first step, whilst external tool calls or actions are left for future work.

\vspace{5pt}

\textbf{LLM call latency and inter-arrival time.}
Inter-arrival time (IAT), the wall-clock gap between consecutive requests, is a foundational metric in traffic characterisation. Alongside burstiness, correlation structure, and flow or request sizes, IAT helps describe whether an arrival process is compatible with common traffic models such as Poisson, heavy-tailed, or self-similar processes, and directly governs queuing behaviour and resource dimensioning. 

In sequential workflows, IAT is strongly coupled to call latency. However, when parallel tasks are introduced, dispatch policy results in several long-running calls having near-zero inter-arrival gaps because they are issued concurrently. Coordination protocols tend to be phase-structured: recruitment, dispatch, discussion, aggregation, and finalization may each generate different request timing patterns. The resulting arrival process may therefore be a mixture of phase-specific processes rather than a single stationary distribution.


\vspace{5pt}
\textbf{Coordination topology.}
Modern multi-agent frameworks make topology an explicit implementation choice. AutoGen supports applications built from multiple conversable agents with programmable conversation behaviours~\cite{wu2023autogen}; AgentVerse structures multi-agent problem solving into staged recruitment, discussion, and consensus phases~\cite{chen2023agentverse}; LangGraph represents agent workflows as graph-structured control flows~\cite{langgraph2026}; and CrewAI supports process-level orchestration of role-specialised agents~\cite{crewai2026}. The topologies these frameworks expose span a range of coordination patterns: a sequential topology issues calls one after another; a star or supervisor-worker topology fans out work from a central coordinator to subagents; and a full mesh allows dense peer-to-peer exchange among all participants. Recent work treats the communication graph as a design variable in its own right: MARBLE studies how topology affects solution quality across different task domains~\cite{zhu2025marble}, while G-Designer optimises communication graphs to reduce unnecessary token overhead~\cite{zhang2024gdesigner}. This body of work has mainly evaluated task quality, robustness, or cost. We instead ask how these topologies affect traffic flow.

\section{Measurement framework}
\label{sec:framework}

The framework measures how agent coordination is expressed in traffic. It observes agentic systems across four layers simultaneously: application, model-serving, container, and network. In the experiments below, topology is the controlled design variable, while the agent runtime, model backend, and monitoring stack are kept unchanged.

\begin{figure}[t]
  \centering
  \includegraphics[width=\columnwidth]{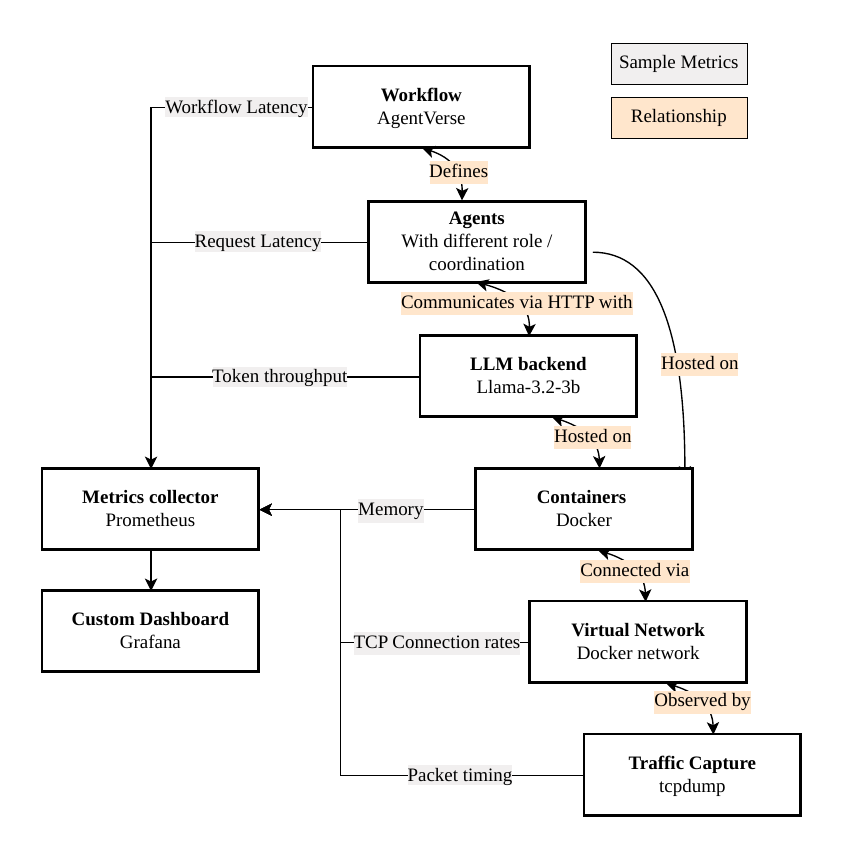}
  \caption{Measurement framework architecture.}
  \label{fig:arch}
\end{figure}

\subsection{Architecture}
\label{sec:arch}

The framework is organised as a set of containerised services connected by Docker bridge networks: the LLM backend, orchestrator, and worker agents each reside on their own dedicated virtual home network, with cross-service traffic flowing over a shared bridge. A passive packet-capture collector reconstructs TCP flows and exports metrics to Prometheus; application traces go to Jaeger; and container resource metrics are collected via cAdvisor. Figure~\ref{fig:arch} sketches the end-to-end architecture.

\subsubsection{Workflow}
\label{sec:arch-workflow}


We use AgentVerse~\cite{chen2023agentverse} as the primary workflow family. AgentVerse structures multi-agent problem solving into staged coordination: agents are recruited for a task, assigned roles, asked to discuss and reach consensus under a selected communication pattern, and then used to produce and evaluate a final answer. The full mesh extension we add covers the densest possible communication case. All experiments are fixed to use four recruited subagents: a mid-range configuration large enough to exhibit fan-out and peer-coordination effects while keeping per-run token volumes within the model's context window.
\begin{figure}[t]
\centering

\captionsetup{type=table}
\footnotesize
\captionof{table}{Metrics collected by the measurement framework, organised by architecture layer.}
\label{tab:metrics}

\resizebox{\columnwidth}{!}{%
\renewcommand{\arraystretch}{0.82}%
\begin{tabular}{p{1.5cm}p{2.2cm}p{3.8cm}}
\toprule
Type & Observed metric & Description \\
\midrule
Network & Traffic volume & Byte and packet counts per directed service pair \\
        & Connection events & Open, clean close, reset, and SYN/SYN-ACK timing events \\
        & Traffic shape & Burstiness, short-window arrivals, and inter-arrival jitter \\
\midrule
Container & Resource usage & Per-container CPU, memory, and interface byte counters \\
\midrule
LLM backend & Inference load & Token throughput, request rate, in-flight requests, and concurrency peaks \\
\midrule
Application & Call latency & End-to-end round-trip time as seen by the agent \\
            & Time-to-first-token & Server-side streaming latency \\
            & Inter-arrival time & Wall-clock gap between consecutive LLM calls within a task \\
            & Token counts & Input and output tokens per call \\
\bottomrule
\end{tabular}%
}

\vspace{15pt}

\captionsetup{type=figure}
\includegraphics[width=\columnwidth]{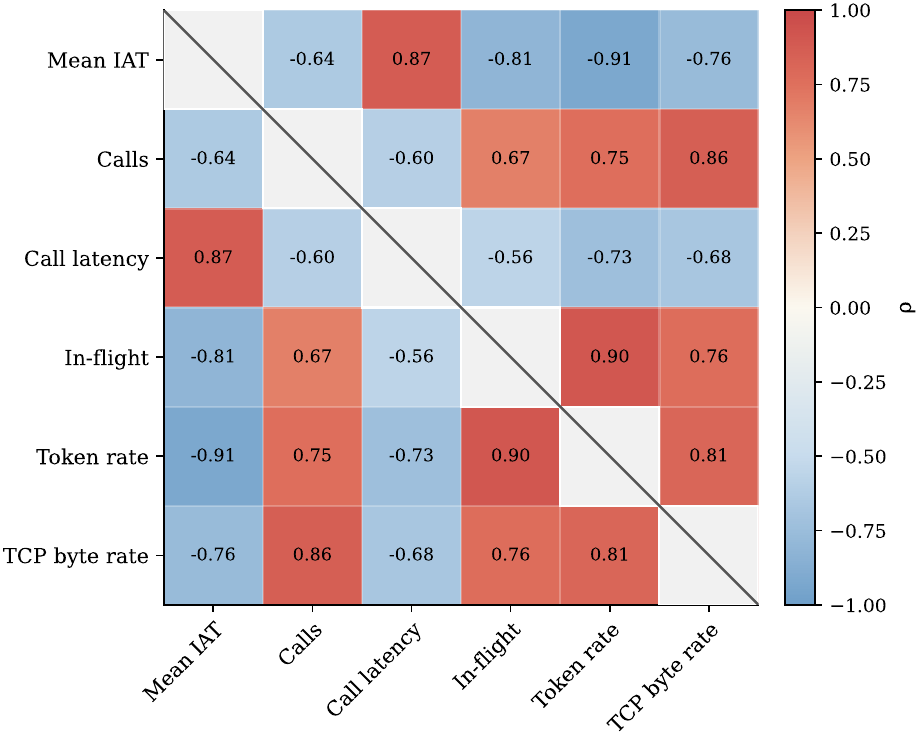}
\captionof{figure}{Spearman rank-correlation matrix over selected per-run metrics pooled across Sequential, Star, and Full Mesh topologies.}
\label{fig:cross-layer-spearman}

\end{figure}

The orchestrator agent drives all discussion-phase LLM call traffic, dispatching calls to worker agents according to the topology's protocol. In the \textbf{sequential} topology agents are called one at a time in a fixed turn order; in the \textbf{star} topology a designated solver proposes first, then all remaining agents are dispatched simultaneously as parallel reviewers; in the \textbf{full mesh} topology all $N{\times}(N{-}1)$ directed peer-message pairs are submitted concurrently each round. Full implementation details, including agent prompts, are available at \url{https://github.com/dlamagna/agentraffic}~\cite{agentraffic2026}.






\begin{figure}[t]
\centering
\footnotesize
\captionof{table}{Discussion-stage metrics for Sequential, Star, and Full Mesh topologies. Percent changes are relative to the sequential baseline.}
\label{tab:sequential-star-metrics}
\resizebox{1.05\columnwidth}{!}{%
\begin{tabular}{lccc}
\toprule
Metric & Sequential & Star & Full Mesh \\
\midrule
Mean LLM requests per repetition & 14.03 & 16.22~\pctup{+16\%} & 29.99~\pctup{+114\%} \\
Total LLM requests & 7{,}013 & 8{,}112 & 14{,}993 \\
Discussion call rate & 0.208~calls/s & 0.209~calls/s~\pctup{+1\%} & 1.366~calls/s~\pctup{+557\%} \\
Duration & 81.1~s & 105.0~s~\pctup{+30\%} & 31.6~s~\pctdown{$-$61\%} \\
Tokens per run & 25.5k & 43.5k~\pctup{+71\%} & 35.2k~\pctup{+38\%} \\
Tokens per call & 1.80k & 2.70k~\pctup{+50\%} & 1.10k~\pctdown{$-$39\%} \\

\midrule
Mean IAT & 4.83~s & 4.87~s~\pctup{+1\%} & 1.09~s~\pctdown{$-$77\%} \\
Median IAT & 4.49~s & 0.40~ms~\pctdown{$-$100\%} & 236~ms~\pctdown{$-$95\%} \\
IAT $p_{95}$ & 8.52~s & 16.20~s~\pctup{+90\%} & 2.82~s~\pctdown{$-$67\%} \\
Burst fraction (IAT $<50$~ms) & 0.0\% & 54.9\% & 38.2\% \\
\midrule
LLM latency $p_{50}$ & 4.89~s & 8.73~s~\pctup{+79\%} & 2.78~s~\pctdown{$-$43\%} \\
LLM latency $p_{95}$ & 10.35~s & 23.94~s~\pctup{+131\%} & 5.10~s~\pctdown{$-$51\%} \\
Time to first token $p_{50}$ & 125~ms & 227~ms~\pctup{+82\%} & 99~ms~\pctdown{$-$21\%} \\
Mean LLM concurrency & 0.97 & 1.62~\pctup{+67\%} & 2.94~\pctup{+203\%} \\
Peak LLM concurrency & 1 & 4~\pctup{+300\%} & 5~\pctup{+400\%} \\
Prompt-token throughput & 237~tok/s & 318~tok/s~\pctup{+34\%} & 1{,}052~tok/s~\pctup{+344\%} \\
Completion-token throughput & 110~tok/s & 178~tok/s~\pctup{+62\%} & 364~tok/s~\pctup{+231\%} \\
\midrule
LLM TCP bytes per run & 50.7k & 85.8k~\pctup{+69\%} & 38.6k~\pctdown{$-$24\%} \\
Mean LLM TCP byte rate & 617~B/s & 768~B/s~\pctup{+25\%} & 1.09k~B/s~\pctup{+77\%} \\
Peak LLM TCP byte rate & 841~B/s & 1.21k~B/s~\pctup{+44\%} & 1.48k~B/s~\pctup{+76\%} \\
\bottomrule
\end{tabular}
}

\vspace{20pt}

\makebox[\columnwidth][c]{%
  \hspace*{0.025\columnwidth}%
  \includegraphics[width=1.1\columnwidth]{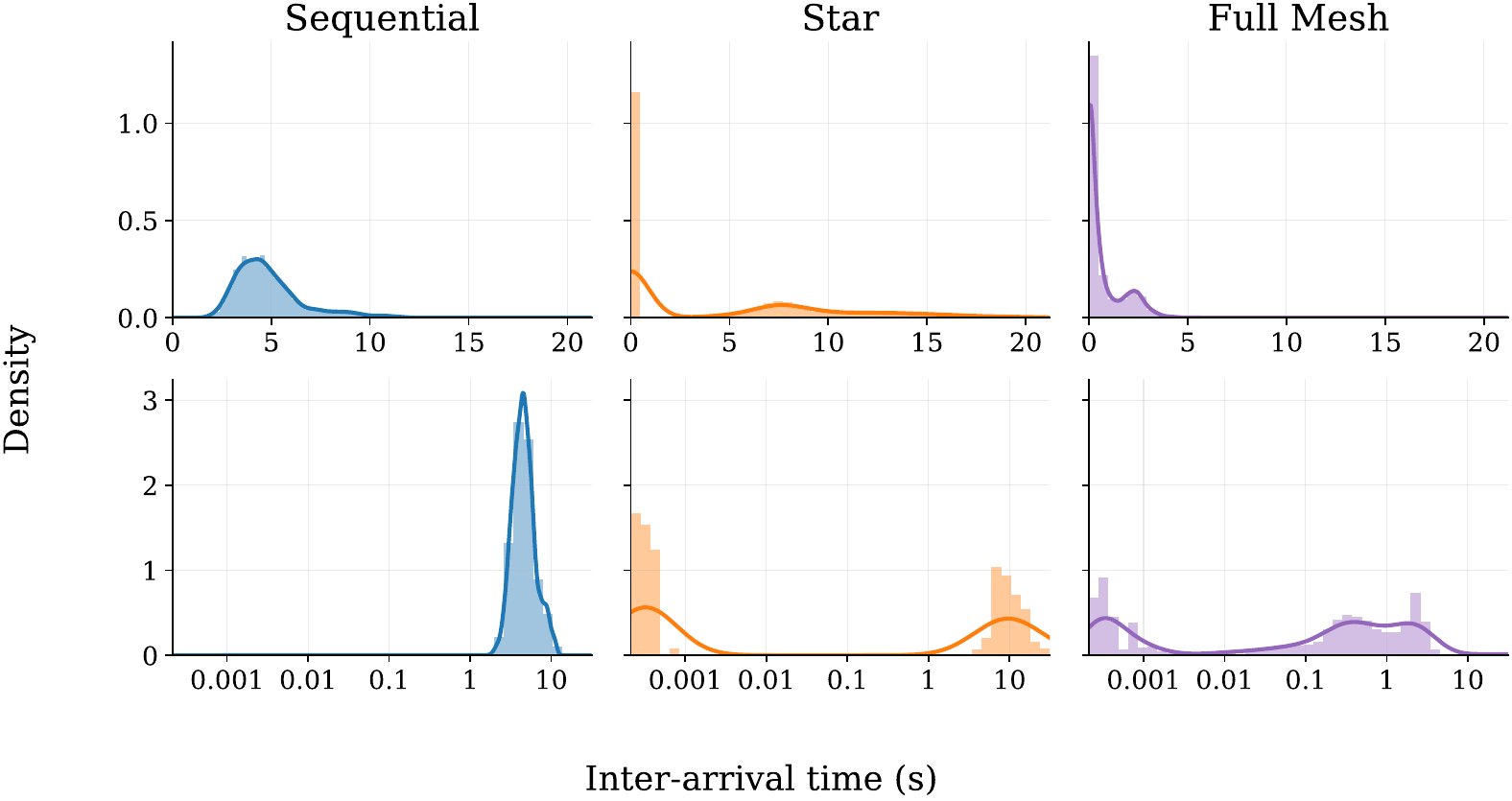}%
}
\vspace{10pt}%
\captionof{figure}{Inter-arrival time distributions for requests to the LLM backend. Rows show linear-scale and log-scale density.}
\label{fig:iat-topologies-combined}
\end{figure}

\subsubsection{LLM backend. }
\label{sec:arch-llm}

All agents use a shared inference backend, which exposes HTTP endpoints for model calls. In the experiments reported in this paper, this service runs an AsyncVLLM backend, implemented using vLLM 0.5's AsyncLLMEngine, with Llama-3.2-3B~\cite{dubey2024llama3}. 
The asynchronous engine accepts concurrent requests from multiple agents and lets vLLM schedule them for batching on the GPU, making queueing, in-flight request count, and token throughput observable parts of the testbed. For this study, request concurrency and token limits were provisioned to avoid scheduler saturation: 
the backend was configured with an effective concurrency limit of eight requests, a maximum model length of 11{,}200 tokens, and a default completion cap of 6{,}144 tokens.


The LLM backend is also where workflow-level context growth can become a serving constraint.
Context accumulation can create prompt-token growth larger than the agent-count increase alone, making token volume a workflow-level mechanism that later appears as model-serving latency, throughput, and traffic-volume effects. Context growth was controlled via token-aware prompt trimming and conciseness instructions to keep request sizes within the model window.



\subsection{Collected metrics}
\label{sec:semcor}

The framework collects metrics across four complementary layers. Most measurements are collected outside the agent execution path: network and container metrics are observed passively, while application-level traces provide semantic context for each run. Table~\ref{tab:metrics} lists all collected metrics.

All layers share the same agent identifiers and timestamps, so per-run metrics can be joined and correlated across layers directly (as in Figure~\ref{fig:cross-layer-spearman}). This multi-layer alignment on a shared time and service-name axis is a key methodological contribution of the framework.

\vspace{5pt}
\textbf{Choice of modelling granularity.}
We model traffic primarily at the LLM-call/request level rather than at the packet level.
In agentic LLM systems, the timing of network load is induced by workflow control flow (agent scheduling, fan-out, and peer coordination).
We therefore treat request-level IATs as the primary workload model and use TCP byte rates, connection events, and packet-level counters as cross-layer evidence that request-level topology effects propagate to underlying infrastructure.







\section{Experiment design and results}
\label{sec:results}

We use IAT (Section~\ref{sec:background}) as the primary metric to test whether common multi-agent coordination patterns induce distinct arrival processes.


We use the AgentVerse workflow family introduced in Section~\ref{sec:arch-workflow} and compare three discussion-stage coordination modes: sequential, star, and full mesh. 
Each topology was repeated 500 times, producing stable IAT distributions.
All runs used the same task source, agent implementation, inference backend, and bridge-level instrumentation. Average values displayed in Table \ref{tab:sequential-star-metrics} are reported from raw samples, whilst the figures and distribution fits use a $3\times$ IQR upper-tail fence to exclude rare extreme gaps that are consistent with model-serving timeouts rather than reasoning phase arrival process.

\subsection{Cross-layer comparison}
\label{sec:comparison}

The shift from sequential to star and full mesh is visible throughout the framework: burst fraction and median IAT change at the application layer, concurrency and token throughput change at the inference layer, and TCP byte rate changes at the network layer (Table \ref{tab:sequential-star-metrics}).

Figure~\ref{fig:cross-layer-spearman} reports Spearman rank correlations over per-run metrics from the same topology experiments. Token throughput, in-flight LLM requests, and TCP byte volume move together, while larger inter-arrival times are negatively associated with backend traffic rate. We read these as consistency checks: the four measurement layers are tracking the same topology-induced changes.

\subsection{Topology effects on IAT}
\label{sec:topology-effects}

Figure~\ref{fig:iat-topologies-combined} shows the three topology distributions. Sequential coordination is the baseline, where recruited agents contribute in strict order. The experiments produced 6{,}513 IAT samples with mean 6.23~s, median 4.56~s, and p95 10.79~s. Figure \ref{fig:iat-topologies-combined} (left) shows it is a homogeneous multi-second arrival process.

Star and full mesh depart sharply from that baseline. Star produces 7{,}355 samples of IAT with mean 6.88~s and p95 20.74~s, but its median collapses to 0.41~ms. 53.3\% of gaps fall below 50~ms, driven by parallel reviewer dispatch. Full mesh is more compressed in time, with 14{,}493 samples, mean 1.09~s, median 236~ms, p95 2.82~s, and 38.2\% of arrivals below 50~ms; its lower per-call token count (1.10k vs 1.80k) reduces individual call latency, while concurrency drives completion-token throughput from 110~tok/s to 364~tok/s. Most notably, the arrival distributions for star and full mesh agentic topologies appear bimodal.

\subsection{Bimodality and distribution shape}
\label{sec:bimodality}

The bimodality in Figure~\ref{fig:iat-topologies-combined} is a structural effect of subagent fan-out, and produces two distinct peaks. The \textbf{fan-out component} (IAT $\leq$ 50~ms) arises when several calls are dispatched almost simultaneously: in a star round, the central solver concurrently triggers parallel reviewer calls; in full mesh, recruited agents send peer messages concurrently at the beginning of a discussion round. The \textbf{reasoning-phase component} (IAT $>$ 50~ms) captures what comes after the initial dispatch, essentially the time spent reasoning by worker agents, and is dictated by model generation time and context accumulation. 
The 50~ms threshold corresponds to the visible trough between modes in Figure \ref{fig:iat-topologies-combined}.

Sequential traffic is approximately unimodal and is the only case where a single log-normal reference is visually plausible. 
Star and full mesh are instead better interpreted as mixture distributions rather than a single arrival process. Figure~\ref{fig:iat-lognormal-fit} adds a log-normal reference fit after filtering the fan-out mode (IAT $\leq$ 50~ms), isolating the reasoning-phase component. Section~\ref{sec:goodness-of-fit} formalises this with distribution fits and model selection.

\subsection{Goodness-of-fit analysis}
\label{sec:goodness-of-fit}

\begin{figure}[t]
  \centering
  \makebox[\columnwidth][c]{%
    \hspace*{-0.05\columnwidth}%
    \includegraphics[width=1.1\columnwidth]{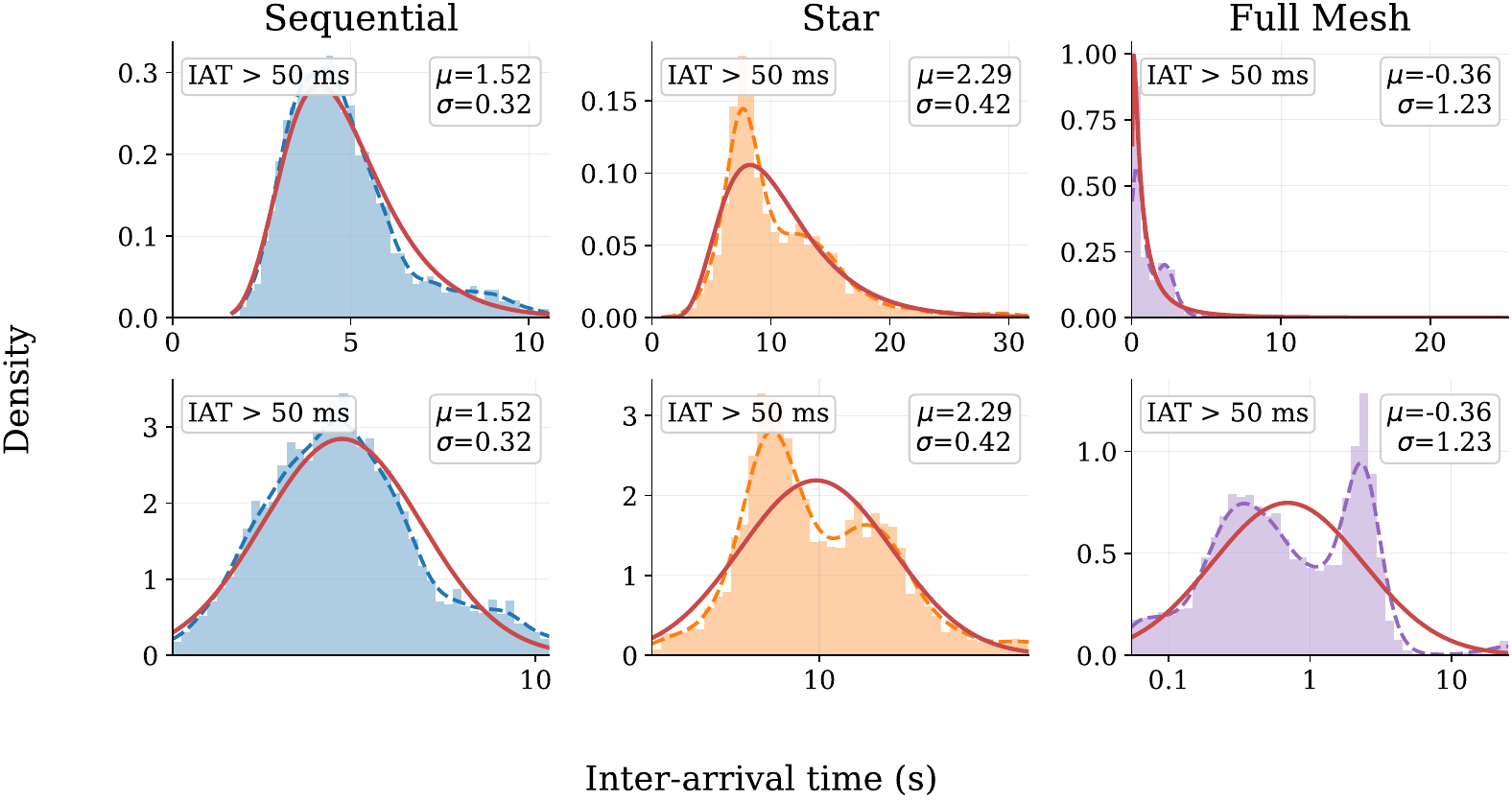}%
  }
  \caption{Log-normal fit of reasoning-phase IATs ($>$ 50~ms). Rows show linear and log scale density.}
  \label{fig:iat-lognormal-fit}
\end{figure}

Star and full mesh produce bimodal IAT distributions (Section~\ref{sec:bimodality}); fitting a single parametric family to the full sample would confuse structurally distinct components.
We therefore restrict fitting to the reasoning-phase component (IAT $>$ 50~ms, Figure~\ref{fig:iat-lognormal-fit}). We fit three candidate families by maximum likelihood: \textbf{Exponential}, the null model for a Poisson (memoryless) arrival process~\cite{paxson1995wide}; \textbf{Weibull}~\cite{weibull1951}, which generalises Exponential with shape parameter $k$
; and \textbf{Log-normal}~\cite{limpert2001lognormal}, which places a Gaussian on log-observations and typically arises from multiplicative processes.

\begin{table}[t]
\centering
\small
\caption{Distribution fits to reasoning-phase IATs (IAT $>$ 50~ms) per topology. Best AIC per topology is \textbf{bold}.}
\label{tab:gof}
\resizebox{\columnwidth}{!}{%
\begin{tabular}{llrccr}
\toprule
Topology & Distribution & $n$ & Parameters & AIC & KS \\
\midrule
Sequential & Exponential & 6{,}168 & mean $=4.77$~s & 31{,}604 & 0.396 \\
           & Weibull     &        & $k=3.09$,\ $\theta=5.32$~s & 23{,}165 & 0.091 \\
           & \textbf{Log-normal} & & $\sigma=0.31$,\ median$=4.54$~s & \textbf{21{,}782} & \textbf{0.027} \\
\midrule
Star       & Exponential & 3{,}303 & mean $=10.55$~s & 22{,}173 & 0.359 \\
           & Weibull     &        & $k=2.38$,\ $\theta=11.92$~s & 19{,}131 & 0.104 \\
           & \textbf{Log-normal} & & $\sigma=0.40$,\ median$=9.70$~s & \textbf{18{,}377} & \textbf{0.077} \\
\midrule
Full Mesh  & Exponential & 8{,}870 & mean $=1.28$~s & 22{,}078 & 0.116 \\
           & Weibull     &        & $k=0.88$,\ $\theta=1.18$~s & 21{,}760 & \textbf{0.068} \\
           & \textbf{Log-normal} & & $\sigma=1.16$,\ median$=0.67$~s & \textbf{20{,}590} & 0.089 \\
\bottomrule
\end{tabular}
}
\begin{flushleft}
{\footnotesize
$\theta$: Weibull scale.\ \ $\sigma$: log-normal shape (std.\ dev.\ in log-space).\ \ median $= e^\mu$ where $\mu$ is the log-space mean.\ \
KS uses test statistic.}
\end{flushleft}
\end{table}

Model selection uses AIC~\cite{akaike1974} and the KS statistic~\cite{massey1951}. At our full sample sizes ($n>3{,}000$) the KS test rejects every candidate~\cite{delbarrio2020}, so the statistic rather than the $p$-value is the meaningful shape-fidelity comparison.
Table~\ref{tab:gof} reports both for each topology, whilst Table~\ref{tab:ks-subsample} shows results from hypothesis testing across different subsample sizes for robustness.

Log-normal achieves the lowest AIC for all three topologies and the smallest KS statistic for sequential and star. However, for full mesh the improvement is not as large, with Weibull achieving a marginally lower KS statistic (0.068 vs.\ 0.089) despite a higher AIC (21{,}760 vs.\ 20{,}590). Since KS is sensitive to the peak whilst AIC penalises overall fit, this suggests Weibull captures the peak more precisely while log-normal better describes the full distribution. 

Subsampling robustness checks confirm the ranking holds at smaller $n$, with log-normal rejected least frequently, indicating the reasoning-phase arrival process is not memoryless.


\begin{table}[!t]
\centering
\footnotesize
\caption{KS rejection rate ($\alpha=0.05$) across 500 random samples of each size, reasoning-phase IATs (IAT $>$ 50~ms).}
\label{tab:ks-subsample}
\resizebox{\columnwidth}{!}{%
\renewcommand{\arraystretch}{0.82}%
\setlength{\tabcolsep}{12pt}%
\begin{tabular}{clccc}
\toprule
$n$ & Distribution & Sequential & Star & Full Mesh \\
\midrule
\multirow{3}{*}{50}  & Exponential & 100\% & 100\% & 51\% \\
                     & Weibull     & 7\%   & 10\%  & 7\%  \\
                     & \textbf{Log-normal}  & 1\%   & 2\%   & 1\%  \\
\midrule
\multirow{3}{*}{100} & Exponential & 100\% & 100\% & 80\% \\
                     & Weibull     & 40\%  & 55\%  & 32\% \\
                     & \textbf{Log-normal}  & 1\%   & 11\%  & 5\%  \\
\midrule
\multirow{3}{*}{200} & Exponential & 100\% & 100\% & 98\% \\
                     & Weibull     & 93\%  & 98\%  & 76\% \\
                     & \textbf{Log-normal}  & 1\%   & 50\%  & 33\% \\
\bottomrule
\end{tabular}%
}
\end{table}

\section{Discussion}
\label{sec:discussion}

The results show that agentic traffic cannot only be described as a response to external user arrivals. When we varied the coordination topology, the structure of the LLM-call arrival process changed with it. The coordination graph acts as part of the workload generator, determining whether agents wait, fan out, or exchange messages concurrently.

The distribution fitting results add precision to this picture. The reasoning-phase gaps are not memoryless: the exponential is decisively rejected across all topologies, and the log-normal's advantage is consistent with call latency arising as a product of prompt length, context accumulation, output token count, and batching decisions. The multiplicative effect of these factors produces a heavy-tailed, non-Poisson inter-arrival process. This is consistent with prior findings in internet traffic measurement, where link traffic volumes have been shown to follow log-normal distributions across a wide range of network conditions and timescales~\cite{alasmar2021}, suggesting that multiplicative combinations of traffic-generating factors consistently produce this distribution shape across different workload classes. Star and full mesh further complicate modelling by mixing this reasoning-phase component with a structural fan-out mode, meaning no single parametric family can effectively describe the full arrival distribution to the LLM backend.

\textbf{Implications.} These findings bring us closer to defining the level of complexity needed to model agentic LLM deployments. Two systems with the same average task rate may place very different demands on the backend if one executes agents serially and another performs coordinated fan-out. Burst-sensitive summaries, including IAT percentiles, sub-second burst fractions, concurrency peaks, and the mixing fraction between fan-out and reasoning-phase arrivals, are useful complements to aggregate throughput. Coordination metadata also matters for interpreting traces: without knowing the workflow topology, a bursty trace may look like load variation when it is a structural feature of the agent protocol.

This connects traffic measurement to current work on multi-agent system design. Prior studies have treated communication topology as a tradeoff between solution quality, robustness, and token cost~\cite{zhang2024gdesigner,zhu2025marble}. The results here suggest that topology also affects the temporal shape of load on shared infrastructure, which is a further dimension alongside model quality and cost that system designers may want to account for. 


\vspace{5pt}

The quantitative values in this paper should be read as controlled-testbed measurements. The experiments use one task family, one agent framework, one model, a single-host deployment, and a single bridge-level capture point; the specific parameter values in Table~\ref{tab:gof} and Table~\ref{tab:sequential-star-metrics} are not intended to generalise beyond this configuration.

\section{Conclusion and future work}
\label{sec:conclusion}

\textbf{Conclusion.}
This paper provided an empirical characterisation of how coordination topology shapes LLM-call arrival processes in multi-agent systems. Across sequential, star, and full-mesh coordination, topology determined whether arrivals formed a homogenous reasoning-phase distribution or a bimodal mixture with a fan-out component. Whilst no single distribution captures the full arrival process, log-normal emerges as the best parametric fit of the reasoning-phase component under all three topologies, with the exponential decisively rejected. Cross-layer measurements confirmed that these structural differences propagate beyond the application layer, affecting inference and network level behaviour.

\vspace{5pt}

\textbf{Future work.} The immediate next step is to broaden the workload matrix. Varying model size, model family, and serving configuration would test how backend characteristics interact with the topology-induced arrival patterns across a wider operating range. 
Adding external tool calls, vector-store queries, or browser actions would introduce additional arrival classes beyond LLM calls. 
A complementary analytical direction is a semi-Markov model of agent execution states, using state-transition probabilities estimated from collected traces to build a principled generative model of the arrival process.

\vspace{5pt}

\textbf{Open-source release.} The framework is open-source at \url{https://github.com/dlamagna/agentraffic}~\cite{agentraffic2026} and is intended as a starting point for measuring agentic traffic under different agent stacks, model backends, and deployment topologies.